\documentclass[prd,twocolumn,showpacs,floatfix,amsmath,nofootinbib,amssymb,floatfix]{revtex4}
\usepackage{graphicx,color,dcolumn,booktabs,bm}
\usepackage{longtable,lscape}
\usepackage{txfonts}
\usepackage{overpic}
\usepackage{amssymb}
\usepackage{indentfirst}
\usepackage{feynmf}   
\usepackage{slashed}  
\usepackage{cases}
\usepackage{color}
\usepackage{multirow}
\usepackage{epstopdf}
\usepackage{graphicx,color,dcolumn,booktabs,bm}
\usepackage[colorlinks, citecolor=blue,anchorcolor=red,menucolor=red, linkcolor=red,filecolor=red,runcolor=red,urlcolor=blue,frenchlinks=red]{hyperref}

\graphicspath{{Figures/}} %

\begin{document}

\title{Implication of the observed $e^+e^-\to p\bar{p}\pi^0$ for studying the $p\bar{p}\to \psi(3770)\pi^0$ process}

\author{Hao Xu$^{1,2}$}\email{xuh2013@lzu.cn}
\author{Ju-Jun Xie$^{1,3,4}$}\email{xiejujun@impcas.ac.cn}
\author{Xiang Liu$^{1,2,4}$\footnote{Corresponding author}}\email{xiangliu@lzu.edu.cn}
\affiliation{$^1$Research Center for Hadron and CSR Physics,
Lanzhou University $\&$ Institute of Modern Physics of CAS,
Lanzhou 730000, China\\
$^2$School of Physical Science and Technology, Lanzhou University,
Lanzhou 730000, China\\
$^3$Institute of Modern Physics, Chinese Academy of Sciences, Lanzhou 730000, China\\
$^4$State Key Laboratory of Theoretical Physics, Institute of Theoretical Physics, Chinese Academy of Sciences, Beijing 100190, China}

\begin{abstract}

We study the charmonium $p \bar{p} \to \psi(3770) \pi^0$ reaction
using effective lagrangian approach where the contributions from
well established $N^*$ states are considered, and all parameters are
fixed in the process of $e^+e^- \to p \bar{p}\pi^0$ at center of
mass energy $\sqrt{s} = 3.773$ GeV. The experimental data on the
line shape of the mass distribution of the $e^+e^- \to
p\bar{p}\pi^0$ can be well reproduced. Based on the studying of
$e^+e^- \to p \bar{p}\pi^0$, the total and differential cross
sections of the $p \bar{p} \to \psi(3770) \pi^0$ reaction are
predicted. At the same time we evaluated also the cross sections of
the $p \bar{p} \to \psi(3686) \pi^0$ reaction. It is shown that the
contribution of nucleon pole to this reaction is largest from
the reaction threshold within a wide range. However, the interference between nucleon
pole and the other nucleon resonance can still change the angle
distributions significantly. Those theoretical results may be test
by the future experiments at $\overline{\mbox{P}}$ANDA.

\end{abstract}


\maketitle

\section{introduction}\label{sec1}

As a forthcoming facility in future, the Anti-Proton Annihilations
at Darmstadt ($\overline{\mbox{P}}$ANDA) experiment will focus on
the production of charmonium, which is govern by nonperturbative
effect of quantum chromodynamics (QCD)~\cite{Lutz:2009ff}. Before
$\overline{\mbox{P}}$ANDA run, there were pioneering theoretical
studies of the charmonium production in the $p\bar{p}$ annihilation
processes~\cite{Gaillard:1982zm,Lundborg:2005am,Barnes:2006ck,Barnes:2007ub,Barnes:2010yb,Lin:2012ru,Pire:2013jva,Wiele:2013vla}.
By calculating two hadron-level diagrams introduced by the Born
approximation, Gaillard and Maiani firstly studied the differential
cross section of the charmonium production plus a soft pion in the
$p \bar{p}$ reaction~\cite{Gaillard:1982zm}. In
Ref.~\cite{Lundborg:2005am}, the cross sections of the chamonium
($\Psi$) production accompanied by a light meson ($m$) from the
process of $p\bar{p} \to \Psi + m$ was calculated by combing with
the measured partial decay widths of charmonium decay into $p
\bar{p} m$. And then, Barnes and Li proposed an initial state light
meson emission model for the near threshold associated charmonium
production processes $p \bar p \to \pi^0 \Psi$
($\Psi=\eta_c,\,J/\psi,\,\psi^\prime,\,\chi_{c0},\,\chi_{c1}$), and
the total and differential cross sections for these reactions were
evaluated~\cite{Barnes:2006ck,Barnes:2007ub,Barnes:2010yb}. It is
also found that the cross section of $p \bar p\to \pi^0 \Psi$ near
threshold may be affected by the Pauli $J/\psi p\bar p$
coupling~\cite{Barnes:2007ub}. Furthermore, Lin, Xu and Liu
revisited the issue of the production of charmonium plus a light
meson at $\overline{\mbox{P}}$ANDA, where the contribution of form
factors (FFs) to these processes are included~\cite{Lin:2012ru}.
Recently, Pire {\it et al.} studied the associated production of a
$J/\psi$ and a pion in antiproton-nucleon annihilation in the
framework of QCD collinear factorization~\cite{Pire:2013jva}, while
in Ref.~\cite{Wiele:2013vla}, the exclusive charmonium production
process $p\bar p\to \pi^0 J/\psi$ was studied within a nucleon-pole
exchange model by including off-shell hadronic FFs and a complete
Lorentz structure with a $\bar p p J/\psi$ Pauli strong coupling.
The contributions from the intermediate $N^*$ states are also
studied in Ref.~\cite{Wiele:2013vla}, and it was found that one can
not ignore the contributions of the $N^*$ resonances in the $\bar{p}
p \to \pi^0 J/\psi$ reaction.

The experimental activity on the charmonium decays have run in
parallel. These decays are of interest because they can be used to
study the associated charomonium production in $p \bar p$
annihilation. In 2014, the BESIII Collaboration reported the
analysis of $e^+e^- \to p \bar p \pi^0$ in the vicinity of $\psi
(3770)$~\cite{Ablikim:2014kxa}. In addition to the Born cross
section of $e^+e^-\to p \bar{p}\pi^0$, the
corresponding $p\pi^0$ and $\bar{p} \pi^0$ invariant mass
distributions of $e^+e^- \to p\bar{p}\pi^0$ process are also
measured~\cite{Ablikim:2014kxa}. These new experimental information in
Ref.~\cite{Ablikim:2014kxa} allows us to further perform a
comprehensive study of $e^+e^- \to \psi(3770) \to p \bar p
\pi^0$, which stimulates our interest to study the contribution of excited nucleon resonances ($N^*$) to $e^+e^- \to \psi(3770)\to
p \bar p \pi^0$ and $\psi(3770)$
production from $p \bar p \to \psi(3770) \pi^0$ reaction.

The nucleon is the simplest system in which the three colors of QCD
can combine to form a colorless object, thus it is important to
understand the internal quark-gluon structure of the nucleon and its
excited $N^*$ states, and the study of excited $N^*$ states is an
interested research field of hadron physics~\cite{Klempt:2009pi},
which can make our knowledge of hadron spectrum abundant. A very
important source of information for the nucleon internal structure
is the $N^*$ mass spectrum as well as its various production and
decay rates, while the charmonium decay into $p \bar{p} \pi^0$ is an
ideal platform to study excited $N^*$ nucleon resonances, because it
provides an effective isospin 1/2 filter for the $\pi N$ system due
to isospin
conservation~\cite{Ablikim:2004ug,Barnes:2011pm,Ablikim:2012zk}.

In this work, we introduce excited $N^*$ nucleon resonances in the
process of $e^+e^- \to \psi(3770) \to p \bar p \pi^0$. By fitting the $p\pi^0$
and $\bar{p}\pi^0$ invariant mass distributions of the cross
section of $e^+e^-\to p\bar{p}\pi^0$, we extract the
information of couplings of $N^* N \pi$ and $\psi(3770) N^*
\bar{N}$, which not only reflects the inner features of discussed
$N^*$, but also helps us to learn the role played by $N^*$ in the
$e^+e^- \to \psi(3770)\to p \bar p \pi^0$.

Based on our studying on the $e^+e^- \to \psi(3770) \to p \bar{p} \pi^0$
process, we move forward to study the $p \bar p \to \psi(3770) \pi^0$
reaction, which is due to the cross relation between the $\psi(3770)
\to p \bar{p} \pi^0$ decay and the $p \bar p\to \psi(3770) \pi^0$
reaction~\cite{Barnes:2011pm}. Here, these extracted parameters from
our study of $e^+e^- \to \psi(3770)\to p \bar p \pi^0$ will be employed to
estimate the production rate of $p \bar p\to \psi(3770)\pi^0$ and
relevant features. We calculate the total and differential cross
sections of the $p \bar{p} \to \psi(3770) \pi^0$ reaction. It is
shown that the contribution of nucleon pole to this reaction is the
largest close to the reaction threshold. However, the interference
between nucleon pole and the other nucleon resonance affects
significantly and could change the angle distributions clearly. Additionally, there were abundant experimental
data of $\psi(3686)\to p\bar{p} \pi^0$ given by
BESIII~\cite{Ablikim:2012zk}, where BESIII released the branching
ratio $B(\psi(3686)\to p \bar p \pi^0) = (1.65 \pm 0.03 \pm 0.15)
\times 10^{-4}$ and the measured $p \pi^0$ and $\bar p \pi^0$
invariant mass spectra~\cite{Ablikim:2012zk}. This experimental
status related to $\psi(3686)$ makes us extend the above study to
the $\psi(3686)\to p \bar{p}\pi^0$ decay, and also the $p \bar p \to
\psi(3686) \pi^0$ reaction. Our
studies provide valuable information to future experimental
exploration of $\psi(3770)$ and $\psi(3686)$ productions plus a pion
through the $p \bar p$ interaction at $\overline{\mbox{P}}$ANDA.

This paper is organized as follows. After introduction in
Sec.~\ref{sec1}, we present the detailed study of $e^+e^- \to
p\bar{p}\pi^0$ by including the excited $N^*$ nucleon resonances
(see Sec.~\ref{sec2}). In Sec.~\ref{sec3}, we further calculate
$p\bar p\to \psi(3770)\pi^0$ by combining with these results
obtained in Sec.~\ref{sec2}. In Sec.~\ref{sec4}, we adopt the
similar approach to study $\psi(3686)\to p\bar{p}\pi^0$ decay and
the $p\bar p\to \psi(3686)\pi^0$ process. The paper ends with a discussion and
conclusion.

\section{Excited $N^*$ nucleon resonance contributions to $e^+e^- \to \psi(3770)\to p\bar{p}\pi^0$} \label{sec2}

\begin{figure}[htpb]
\begin{center}
\includegraphics[scale=0.5]{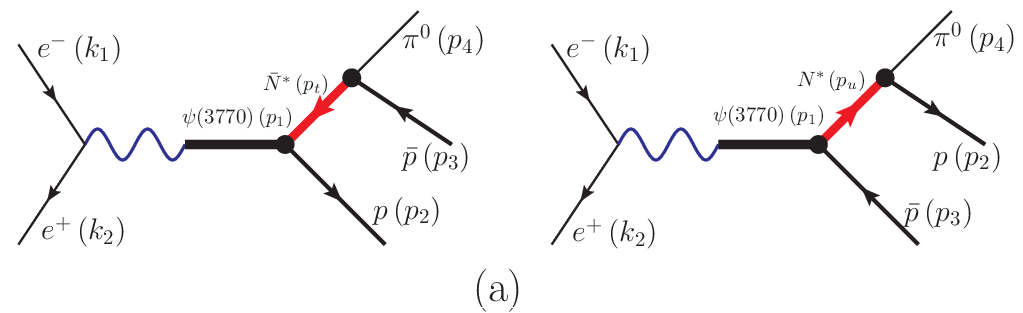}
\includegraphics[scale=0.5]{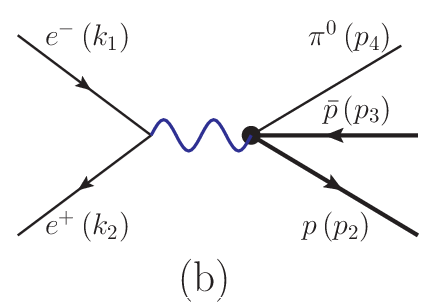}
\caption{(color online). The Feynman diagrams for the process $e^+e^-  \to p \bar{p} \pi^0$ in the vicinity of $\psi(3770)$.}\label{dec}
\end{center}
\end{figure}

First, we study the process $e^+e^-  \to p \bar{p} \pi^0$ with an
effective Lagrangian approach. In hadron level, the process $e^+e^-
\to p \bar{p} \pi^0$ in the vicinity of $\psi(3770)$ is described by
the diagrams shown in Fig.~\ref{dec}. In Fig.~\ref{dec} (a), $e^+$
and $e^-$ annihilate into photon, which couples with charmonium
$\psi(3770)$. And then, $\psi(3770)$ interacts with final states,
where we consider the contributions from nucleon-pole ($ \equiv
P_{11}$) with $J^{P} = \frac{1}{2}^+$ and five $N^*$ states that are
well established~\cite{Agashe:2014kda}: $N(1440)$ ($ \equiv P_{11}$)
with $J^{P}=\frac{1}{2}^+$, $N(1520)$ ($ \equiv D_{13}$) with
$J^{P}=\frac{3}{2}^-$, $N(1535)$ ($ \equiv S_{11}$) with
$J^{P}=\frac{1}{2}^-$, $N(1650)$ ($ \equiv S_{11}$) with
$J^{P}=\frac{1}{2}^-$, and $N(1720)$ ($ \equiv P_{13}$) with
$J^{P}=\frac{3}{2}^+$. Additionally, we also consider the background
contribution, where the $e^+e^-$ annihilation directly into $p
\bar{p} \pi$ without intermediate $\psi(3770)$, which is shown in
Fig.~\ref{dec} (b).

To compute the contributions of these terms, we use the effective
interaction Lagrangian densities for each vertex. For the $\gamma \psi(3770)$ coupling, we adopt the vector meson dominant (VMD) model, where a vector meson couples to a photon is described by \cite{Lin:2013mka}
\begin{eqnarray}
\mathcal {L}_{V\gamma} = - \frac{e M^2_V }{f_V} V_\mu A^\mu.
\end{eqnarray}
In above expression, $M_V$ and $f_V$ are the mass and the decay
constant of the vector meson, respectively. The decay constant
$e/f_V$ can be fitted through $V \rightarrow e^+ e^-$:
\begin{eqnarray}
e/f_V = \left[\frac{ 3 \Gamma_{V \rightarrow e^+ e^- } M_V^2} { 8 \alpha |\vec{p}|^3}\right]^{1/2} \simeq \left[\frac{3\Gamma_{V\to e^+ e^-}}{\alpha M_V}\right]^{1/2} \ ,
\end{eqnarray}
where $|\vec{p}|=(M_V^2-4m_e^2)^{1/2}/2\simeq M_V/2$ is three momentum of an electron in the rest
frame of the vector meson. $\alpha = e^2/(4\pi) = 1/137$.
Using $B(\psi(3770)\to e^+e^-)=(9.6\pm 0.7)\times 10^{-6}$~\cite{Agashe:2014kda}, we obtain $e/f_{\psi(3770)}= 0.0053$.

The $J/\psi N \bar{N}$ and $N N\pi$ couplings are described by:
\begin{eqnarray}\label{npi}
\mathcal {L}_{\pi NN}&=&-\frac{g_{\pi NN}}{2m_N}\bar{N}\gamma_5\gamma_\mu\tau\cdot\partial^\mu\pi N , \label{pinn} \\
\mathcal {L}_{\psi NN}&=&-g_{\psi NN}\bar{N}\gamma_\mu V^\mu N,
\label{vnn}
\end{eqnarray}
where $V^\mu$ stands for the vector field of $\psi(3770)$. We
take $g_{\pi NN} = 13.45$.

For the $N^* N \pi$ and $\psi N^* \bar N$ vertexes, we adopt the
Lagrangian densities as used in
Refs.~\cite{Tsushima:1996xc,Tsushima:1998jz,Ouyang:2009kv,Wu:2009md,Cao:2010km,Cao:2010ji,Zou:2002yy}:
\begin{eqnarray}
\mathcal {L}_{\pi NP_{11}}&=&-\frac{g_{\pi NP_{11}}}{2m_N}\bar{N}\gamma_5\gamma_\mu\tau\cdot\partial^\mu\pi R_{P_{11}}+h.c. ,\label{nspi}  \\
\mathcal {L}_{\pi NS_{11}}&=&-g_{\pi NS_{11}}\bar{N}\tau\cdot\pi R_{S_{11}}+h.c.,  \\
\mathcal {L}_{\pi NP_{13}}&=&-\frac{g_{\pi NP_{13}}}{m_N}\bar{N}\tau\cdot\partial_\mu\pi R^\mu_{P_{13}}+h.c.,  \\
\mathcal {L}_{\pi ND_{13}}&=&-\frac{g_{\pi
ND_{13}}}{m_N^2}\bar{N}\gamma_5\gamma^\mu\tau\cdot\partial_\mu\partial_\nu\pi
R^\nu_{D_{13}}+h.c., \\
\mathcal {L}_{\psi NP_{11}}&=&-g_{\psi NP_{11}}\bar{N}\gamma_\mu V^\mu R_{P_{11}}+h.c. ,  \\
\mathcal {L}_{\psi NS_{11}}&=&-g_{\psi NS_{11}}\bar{N}\gamma_5 \gamma_\mu V^\mu R_{S_{11}}+h.c.,  \\
\mathcal {L}_{\psi NP_{13}}&=&-ig_{\psi NP_{13}}\bar{N}\gamma_5 V_\mu R^\mu_{P_{13}}+h.c.,  \\
\mathcal {L}_{\psi ND_{13}}&=&-g_{\psi ND_{13}}\bar{N}V_\mu
R^\mu_{D_{13}}+h.c. ,\label{nsv}
\end{eqnarray}
where $R$ is a $N^*$ field.

For the intermediate nucleon-pole or $N^*$ state, a Breit-Wigner
form of its propagator $G_J (q)$ can be written
as~\cite{Huang:2005js}
\begin{eqnarray} \label{p12}
G_{\frac{1}{2}}(q)=i\frac{\slashed{q}+M_{N^*}}{q^2-M^2_{N^*}+iM_{N^*}
\Gamma_{N^*} }
\end{eqnarray}
for $J = \frac{1}{2}$, and
\begin{eqnarray} \label{p32}
G^{\mu\nu}_{\frac{3}{2}}(q) &=& i\frac{\slashed{q}+M_{N^*}}{q^2-M^2_{N^*}+iM_{N^*}\Gamma_{N^*} }\Bigg(-g_{\mu\nu}+\frac{1}{3}\gamma_\mu\gamma_\nu \nonumber \\
& &+\frac{1}{3m}(\gamma_\mu q_\nu-\gamma_\nu
q_\mu)+\frac{2}{3}\frac{q_\mu q_\nu}{q^2}\Bigg)
\end{eqnarray}
for $J=\frac{3}{2}$. In Eqs.~(\ref{p12}) and (\ref{p32}), $M_{N^*}$
and $\Gamma_{N^*}$ are the masses and widths of these intermediate
$N^*$ states, respectively. The values used in the present work for
$M_{N^*}$ and $\Gamma_{N^*}$ are summarized in
Table.~\ref{table:Nstar}.

\renewcommand{\arraystretch}{1.5}
\begin{table}[htbp]
\caption{Relevant resonant parameters for $N^*$ states. The values are taken
from Particle Data Book~\cite{Agashe:2014kda}.}
\label{table:Nstar}
\begin{center}
\begin{tabular}{lcc} \toprule[0.6pt]\toprule[0.6pt]
            $N^*$     & $M_{N^*}$ (MeV)    & $\Gamma_{N^*}$ (MeV)    \\ \midrule[0.2pt]
            $N(938)$  & $938$   & $0$     \\ 
            $N(1440)$ & $1430$  & $350$     \\ 
            $N(1520)$ & $1515$  & $115$   \\ 
            $N(1535)$ & $1535$  & $150$  \\ 
            $N(1650)$ & $1655$  & $140$  \\
            $N(1720)$ & $1720$  & $250$  \\  \bottomrule[0.7pt]\bottomrule[0.7pt]
        \end{tabular}
    \end{center}
\end{table}

On the other hand, we also need to introduce the form factors for these
intermediate off-shell $N^*$ ($N$), which are taken as in
Refs.~\cite{Feuster:1997pq,Haberzettl:1998eq,Yoshimoto:1999dr,Oh:2000zi}:
\begin{eqnarray}\label{ff}
F(q^2)=\frac{\Lambda^4}{\Lambda^4 + (q^2-M_{N^*}^2)^2},
\end{eqnarray}
where the cutoff parameter $\Lambda$ can be parameterized as
\begin{eqnarray}\label{cut}
\Lambda = M_{N^*} + \beta \Lambda_{QCD},
\end{eqnarray}
with $\Lambda_{QCD} = 220$ MeV. The parameter $\beta$ will be
determined by fitting the experimental data.

For the background contribution depicted in Fig.~\ref{dec} (b), we construct the amplitude in analogy of Ref.~\cite{Chen:2010nv}:
\begin{eqnarray} \label{directamplitude}
\mathcal{M}_{NoR}=g_{NoR} \bar{v}(k_2) e \gamma^\mu u(k_1)
\frac{1}{s} \bar{u}(p_2) \gamma^\mu \gamma_5 v(p_3)
\mathcal{F}_{NoR}(s),
\end{eqnarray}
with $\mathcal{F}_{NoR}(s) = \textnormal{exp} ( -a ( \sqrt{s} -
\sum_f m_f )^2)$, where $ \sum_f m_f$ means the mass of the final
states are summed over. The parameter $a$ will be fitted to the
experimental measurements, and $s$ is the invariant mass square of
the $e^+ e^-$ system.

In the phenomenological Lagrangian approaches, the relative phases
between amplitudes from different diagrams are not fixed. Generally,
we should introduce a relative phase between different amplitudes as
free parameters, and the total amplitude can be written as:
\begin{eqnarray} \label{mpppi}
&&\mathcal{M}_{e^+ e^- \rightarrow p \bar{p} \pi^0} \nonumber \\
 &=& \mathcal{M}_{NoR}e^{i\phi_{NoR}} + \bar{v}(k_2) e \gamma^\mu u(k_1) \frac{-g^{\mu\nu}}{s} e m^2_{\psi} /f_{\psi} \nonumber \\
    && \times \frac{ -g_{\nu\alpha}+\frac{p_{\psi\nu} p_{\psi\alpha}}{m^2_\psi} } {s-m^2_\psi+im_\psi\Gamma_\psi}\left(\mathcal{M}_{N}^\alpha+\sum_{N^*}\mathcal{M}_{N^*}^\alpha e^{i\phi_{N^*}}\right) ,
\end{eqnarray}
where $\mathcal{M}_{N^*(N)}^\alpha$ describing the subprocesses $\psi(3770) \to p\bar{p}\pi^0$ are given completely in appendix.

The differential cross section is given by~\cite{Xie:2015zga}
\begin{eqnarray} \label{gpppi}
d\sigma_{e^+ e^- \rightarrow p \bar{p} \pi^0}=\frac{(2\pi)^4 \sum |\mathcal{M}_{e^+ e^- \rightarrow p \bar{p} \pi^0}|^2}{4\sqrt{(k_1 \cdot k_2)^2}}d\Phi_3,
\end{eqnarray}
and the phase space factor is given by
\begin{eqnarray}
d\Phi_3=\frac{1}{(2\pi)^9}\frac{1}{8\sqrt{s}}|\vec p^*_3||\vec
p_2|d\Omega^*_3 d\Omega_2 dm_{\bar{p}\pi},
\end{eqnarray}
with $\sum |{\cal M}|^2$ averaging over the spins of the initial
$e^+e^-$ and summing over the polarizations of the final states $p
\bar{p}$.

As we can see in the appendix, in the tree-level approximation, only
the products like $g_{N^*} \equiv g_{V N N^*}g_{\pi N N^*}$ enter in
the invariant amplitudes. They are determined with the use of
MINUIT, by fitting to the low energy experimental data on mass
distribution of $e^+e^- \to p\bar{p} \pi^0$ at $\sqrt{s} = 3.773$
GeV~\cite{Ablikim:2014kxa}. So far we have fifteen unknown
parameters: six $g_{N^*}$, six phase angles $\phi_{N^*}$ and
$\phi_{NoR}$, one cutoff $\beta$ in the form factors and two
parameters $g_{NoR}$ and $a$ in direct production amplitude
Eq.~(\ref{directamplitude}). We perform those fifteen-parameter
$\chi^2$ fits to the BESIII experiment data on the invariant mass
distribution at $3.773$ GeV below $1.8$ GeV, and make use of the
total cross section information in Ref.~\cite{Ablikim:2014kxa}.
Here, we do not consider the invariant mass region beyond $1.8$ GeV,
where contains large contribution from higher mass $N^*$ states and
other complicated resonance which decays to $p\bar{p}$. In
Ref.~\cite{Wiele:2013vla}, it was pointed that in the case of
$\bar{p} p \to \pi^0 J/\psi$ reaction the higher mass $N^*$
resonances are needed. Indeed, in the present case, if we go beyond
$1.8$ GeV, we need also the higher mass $N^*$ states. On the other
hand, we did also another calculation including the contributions of
higher spin nuclear excited states, $N(1675)5/2^-$ and
$N(1680)5/2^+$. It is find that their contributions are quite small
and the fitted parameters for the other nuclear resonance are little
changed. Thus, we will not include the contributions of this two
states in this work.

\begin{figure}[htpb] \centering
\includegraphics[scale=0.3]{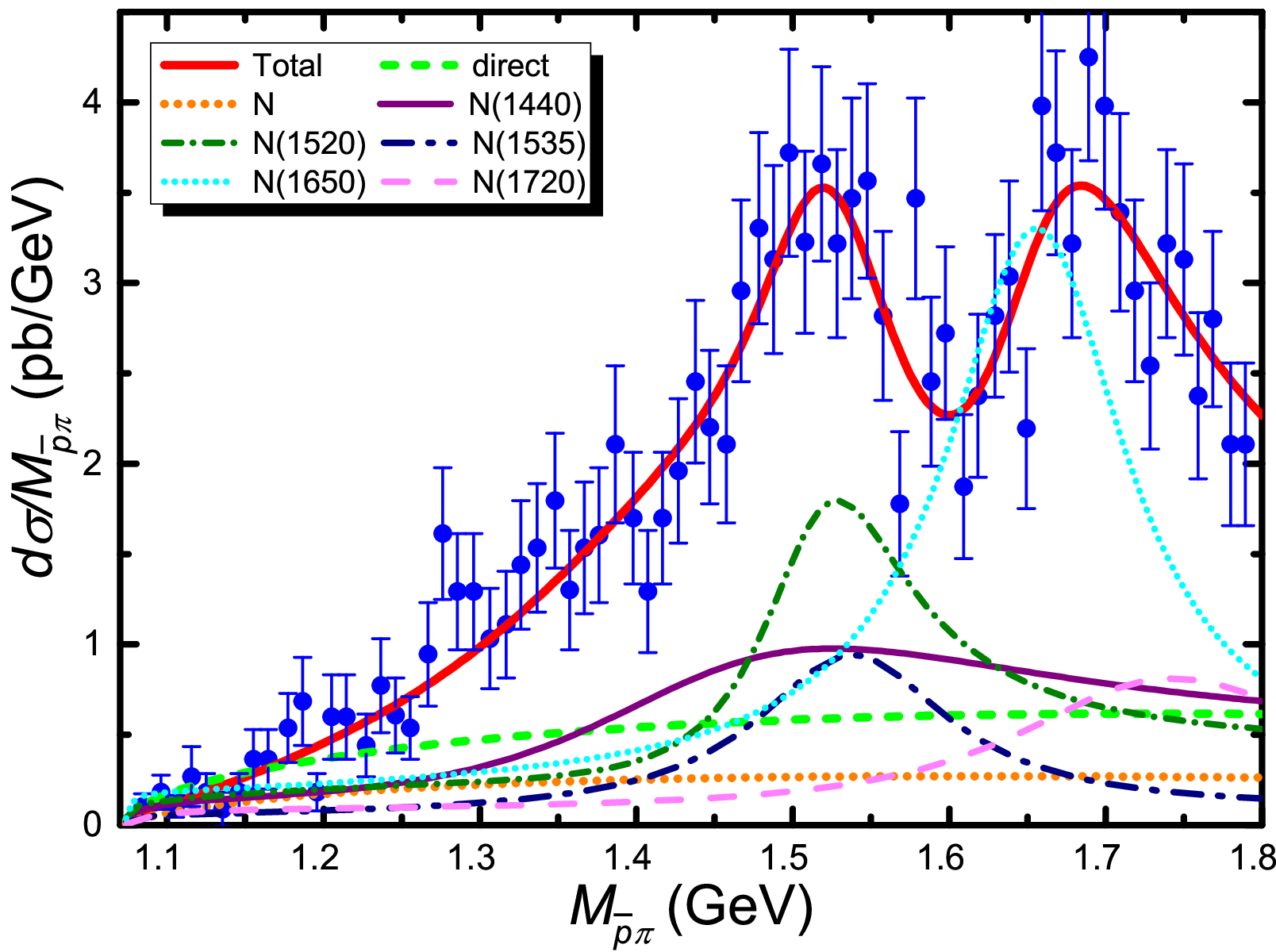}
\caption{(color online). The fitted mass spectrum of the process
$e^+e^-\rightarrow  p\bar{p}\pi^0$ at $\sqrt{s} = 3.773$ GeV
comparing to the experiment data. The experiment data are taken from
Ref.~\cite{Ablikim:2014kxa}. The green dashed line stands for the
background contribution, the orange doted line stands for the
nucleon-pole contribution, the red line is the full result, and
other lines show the contributions from different $N^*$ resonances.
Notice that the experimental event is converted to physical
differential cross section using the experimental total cross
section at $\sqrt{s} = 3.773$
GeV~\cite{Ablikim:2014kxa}.}\label{figdec2}
\end{figure}

We get a minimal $\chi^2/dof = 1.03$ with the fitted cut-off
parameter $\beta=5.0\pm2.1$. The parameters appearing in direct
amplitude Eq.~(\ref{directamplitude}) are $g_{\textnormal{NoR}}=0.62\pm0.09$, $\phi_{\textnormal{NoR}}=0.96\pm0.51$ rad and $a=0.81\pm0.03$ GeV$^{-2}$.
The other fitted
parameters are compiled in Table~\ref{ngphi}. The fitted results are
shown in Fig.~\ref{figdec2} compared with the experimental data
taken from Ref.~\cite{Ablikim:2014kxa}, where the green dashed line
stands for the background contribution, the orange doted line stands
for the nucleon-pole contribution, the red line is the full
result, and other lines show the contributions from different $N^*$
resonances. Notice that we have converted the experimental event to
physical differential cross section using the experimental value
$\sigma_{total}=7.71$ pb at $3.773$ GeV \cite{Ablikim:2014kxa}. Our
results can describe the two clear peaks around 1.5 GeV and 1.7 GeV,
thanks to the contributions from $N(1520)$, $N(1535)$ and $N(1650)$
resonances. The contribution from the nucleon pole is small, while
the background contribution is relatively large.

In Fig.~\ref{figdec2}, it is interesting to see large interfering
effects between different contributions. At the low $M_{\bar{p}\pi}$
region around $1.1-1.3$ GeV, large cancelation between the nucleon
pole, $N^*$ and the background leads to quiet suppressed spectrum. From the
two-peak region around $1.4-1.8$ GeV, we can directly see that, the
background contribution plus $N^*$ contribution (means without
interfering contribution) is far beyond the dip at 1.6 GeV, it
indicates a large cancelation between the background contribution
and $N^*$ contributions thanks to the interfering effect.

\setlength{\tabcolsep}{21pt}
\renewcommand{\arraystretch}{1.5}
\begin{table}[htbp]
	\caption{The fitted parameters in the process $e^+e^- \to p
		\bar{p}\pi^0$, where $g_{N^*}=g_{\psi(3770)NN^*}g_{\pi N N^*}$. For nucleon, $g_{N}$ 
		is defined as $g_{N}=g_{\psi(3770)NN}g_{\pi N N}$ . }
	\label{ngphi}
	\begin{center}
		\begin{tabular}{lcc} \toprule[0.6pt]\toprule[0.6pt]
			$N^*$     & $g_{N^*}\,(\times10^{-3})$    & $\phi_{N^*} \,(\textnormal{rad})$    \\ \midrule[0.2pt]
			$N(938)$ & $66.36 \pm 3.54$   & $--$   \\ 
			$N(1440)$ & $36.96 \pm 0.07$   & $2.83\pm0.04$   \\ 
			$N(1520)$ & $34.14 \pm 4.31$   & $1.92\pm0.69$    \\ 
			$N(1535)$ & $5.86 \pm 1.91$   & $0.82\pm0.54$    \\ 
			$N(1650)$ & $10.34 \pm 2.09$   & $5.23\pm0.64$   \\
			$N(1720)$ & $17.35 \pm 6.65$   & $0.12\pm0.83$    \\  \bottomrule[0.7pt]\bottomrule[0.7pt]
		\end{tabular}
	\end{center}
\end{table}

\section{The $\psi(3770)$ production  in the process $p\bar p\to \psi(3770)\pi^0$}\label{sec3}

\begin{figure}[htpb]
\centering
\includegraphics[scale=0.5]{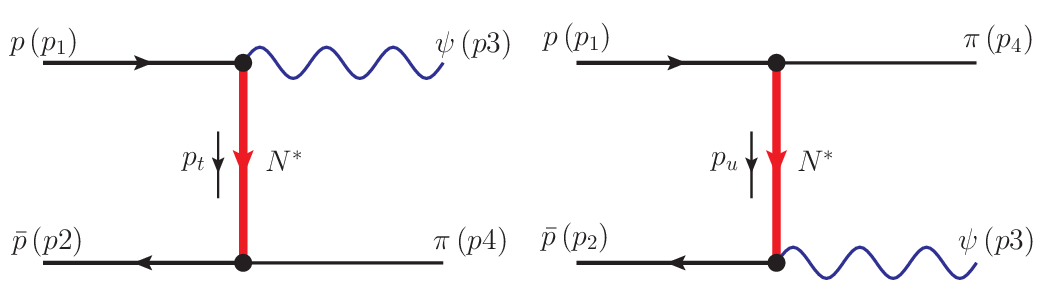}
\caption{(color online). The typical Feynman diagrams for the
process $p \bar p \to \pi^0 \psi(3770)$.}\label{cro}
\end{figure}

A charmonium plus a light meson $\pi$ can produced by the low energy
$p\bar p$ annihilation process. The tree level diagrams for the
$p\bar p \to \psi(3770) \pi^0$ reaction are depicted in
Fig.~\ref{cro}. It is worth to mention that the effect of the $N^*$
resonances in the cross channel of Fig.~\ref{cro} has been studied
firstly in the $\bar{p} p \to \pi^0 J/\psi$
reaction~\cite{Wiele:2013vla}. It was found that the contributions
from the $N^*$ resonances in the $\bar{p} p \to \pi^0 J/\psi$
reaction are important. In the present work, we extend the model of
Ref.~\cite{Wiele:2013vla} to the process of the higher charmonium
states [$\psi(3770)$ and $\psi(3686)$] production.~\footnote{We
mention that the Regge exchange may be important, unfortunately, the
information of the Regge propagators are scarce and we hope we can
include the Regge contribution in the future.}

The differential cross section of the $p \bar p \rightarrow \pi^0
\psi(3770)$ reaction at center of mass (c.m.) frame can be expressed
as~\cite{Agashe:2014kda}
\begin{eqnarray} \label{gcro}
\frac{d\sigma_{p \bar p \to \pi^0  \psi(3770)}}{dcos\theta}
=\frac{1}{32 \pi s}\frac{|\vec{p}^{~\rm cm}_{3}|}{|\vec{p}^{~\rm
cm}_{1}|} \sum \overline{|\mathcal{M}|^2},
\end{eqnarray}
where $\theta$ denotes the angle of the outgoing $\pi^0$ relative to
beam direction in the $\rm c.m.$ frame, $\vec{p}^{~\rm cm}_{1}$ and
$\vec{p}^{~\rm cm}_3$ are the three-momentum of the proton and
$\psi(3770)$ in c.m. frame, respectively, while the total invariant
scattering amplitude $\cal M$ is given in appendix using
cross symmetry.

\begin{figure}[htpb]
\centering
\includegraphics[scale=0.3]{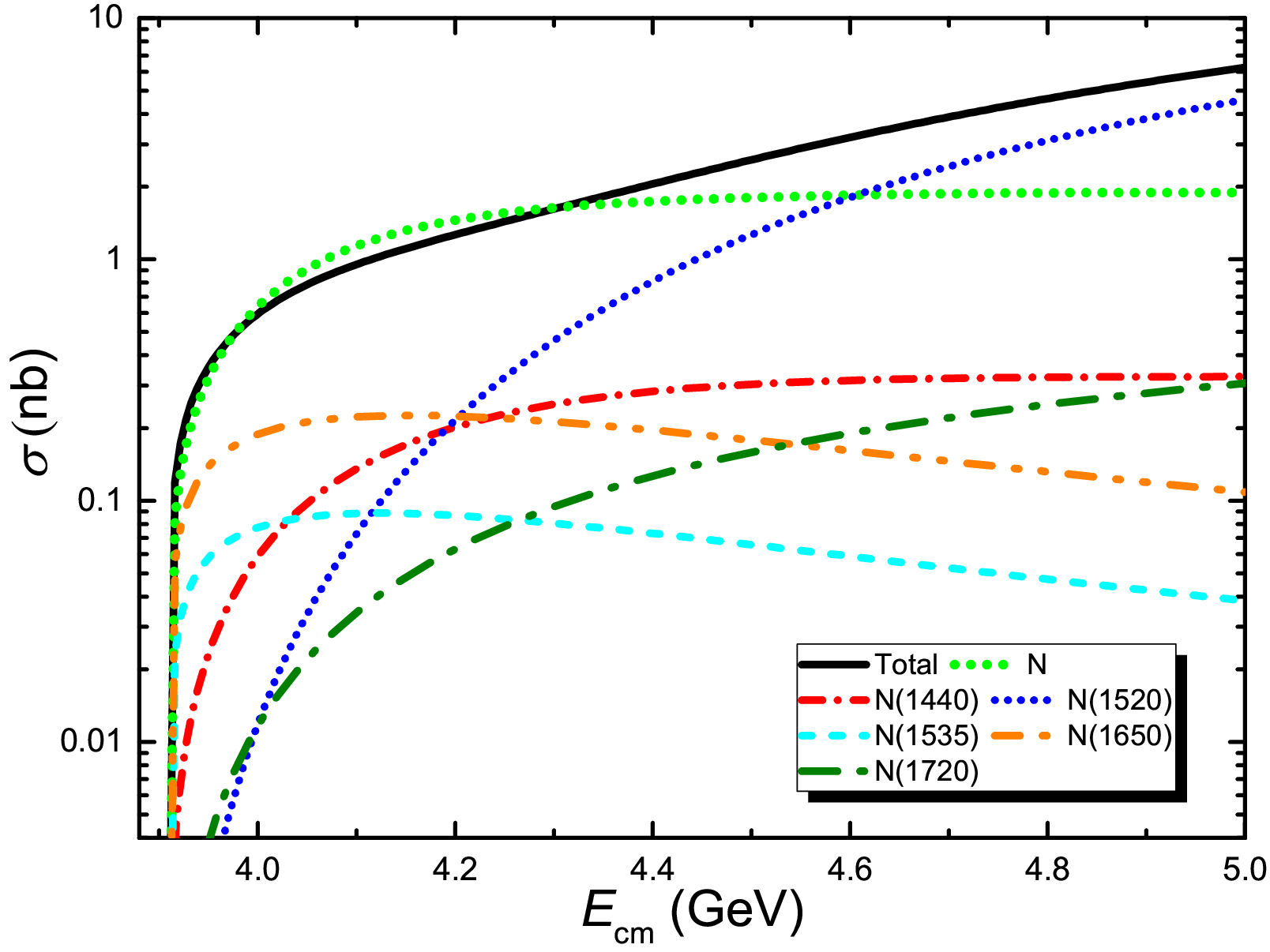}
\caption{(color online). Total cross section of the $p \bar p \to
\pi^0 \psi(3770)$ reaction. The black line is total result, and
other lines show the contributions from different $N^*$
resonances.}\label{figcros21}
\end{figure}

With the parameters determined from the process of $e^+e^- \to p
\bar p \pi^0$, we calculate the total and differential cross
sections of $p\bar{p} \to \pi^0 \psi(3770)$ reaction. In
Fig.~\ref{figcros21}, we show our results for the total cross
section of the $p\bar{p} \to \pi^0 \psi(3770)$ reaction as a
function of the invariant mass ($E_{\rm cm}$) of $\bar p p$ system.
At $E_{\rm cm} = 5.26$ GeV, the total cross section is $8.25$ nb,
however it is beyond the upper limit of the value obtained in
Ref.~\cite{Ablikim:2014kxa}.

From Fig.~\ref{figcros21}, we see that the nucleon pole gives
largest contribution, and becomes dominant in the region $E_{\rm cm}
< 4.6$ GeV. This is because in the reaction of $ p \bar p \to
\psi(3770) \pi^0$, the four momentum square, $q^2$, of nucleon or
other nucleon resonance is smaller than 0, and the propagator
$\frac{1}{q^2 - M^2}$ will increase the contribution of nucleon
because of its small mass. Besides, it is found that the
contributions from $N^*$ state with different quantum numbers have
quite different behavior. The contributions from $N(1535)$ and
$N(1650)$ with $J^{P} = \frac{1}{2}^-$ decrease at $E_{\rm cm}$
around $4.2$ GeV, while the others increase all the time. Overall,
the total cross section always increases.

\begin{figure}[htpb]
\centering
\includegraphics[scale=0.3]{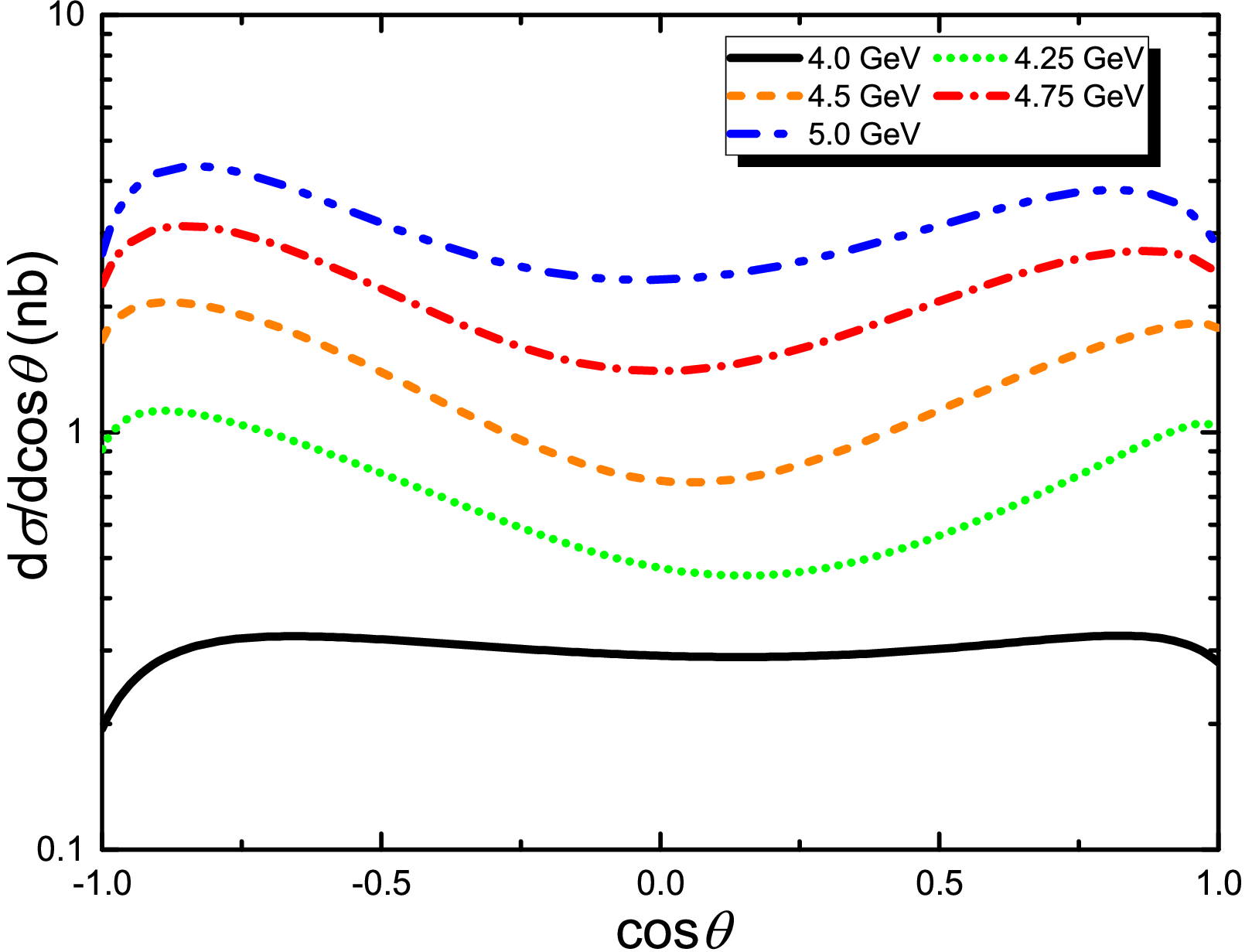}
\caption{(color online). Angular distributions of the $p \bar p
\rightarrow \pi^0 \psi(3770)$ reaction with full
contribution.}\label{figangle1}
\end{figure}

In addition, we also calculate the angular distribution of the $p
\bar p \rightarrow \pi^0  \psi(3770)$ reaction at $E_{\rm cm} =
4.0$, $4.25$, $4.5$, $4.75$ and $5.0$ GeV. The numerical results are
shown in Fig.~\ref{figangle1}. We can see that there emerges an
obvious peak at the backward angles (around ${\rm cos}\theta \sim
0.8$) at $E_{\rm cm} \ge 4.25$ GeV produced by the contributions of
nucleon results in the $u$-channel, while the larger results at the
forward angles is due to the $t$-channel nucleon resonances
contributions.

\begin{figure}[htpb]
\centering
\includegraphics[scale=0.3]{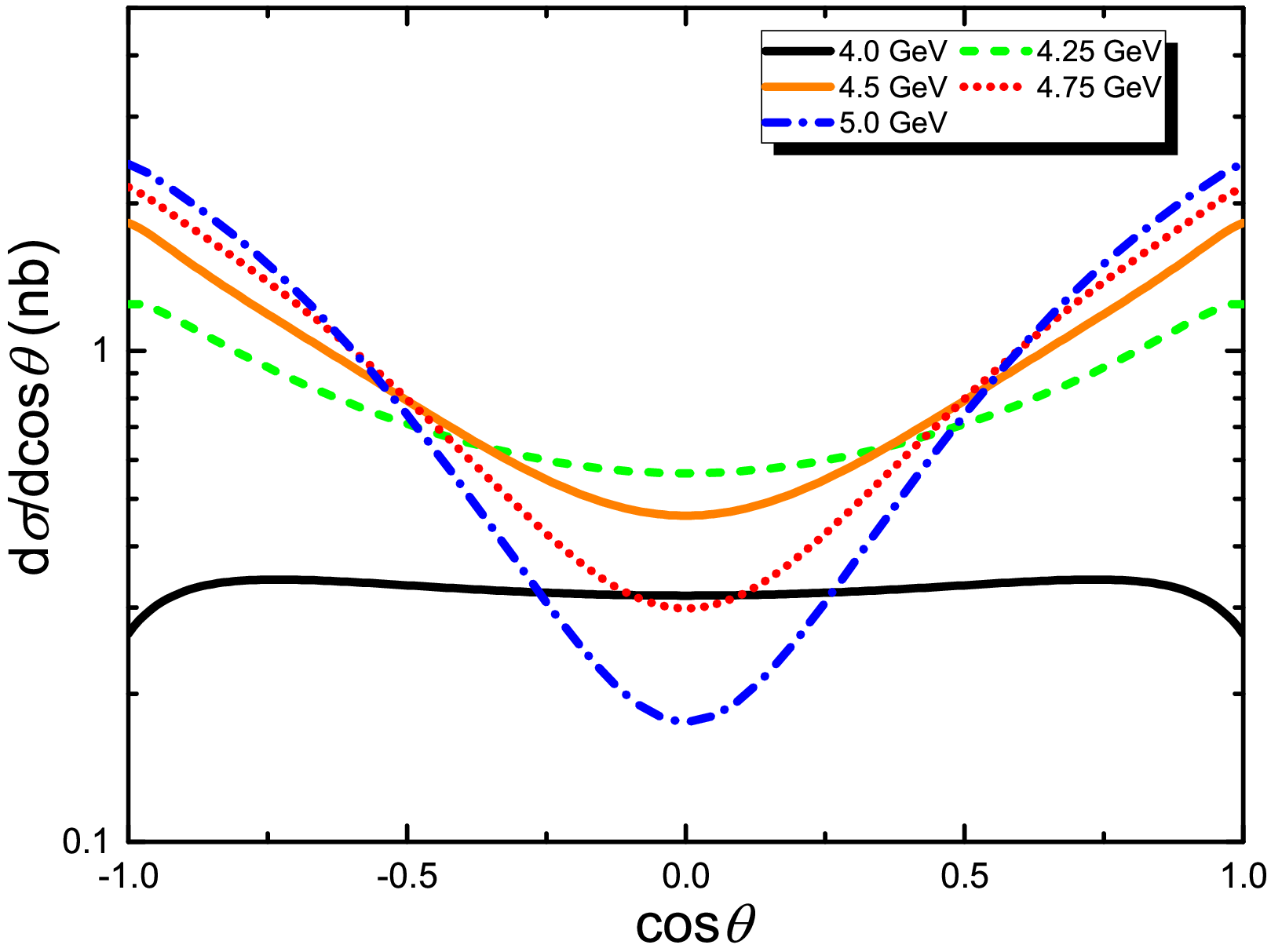}
\caption{(color online). Angular distributions of the $p \bar p
\rightarrow \pi^0 \psi(3770)$ reaction considered only the
contribution from the nucleon pole.}\label{figangle1n}
\end{figure}

In Fig.~\ref{figangle1n}, we show the numerical results of the
angular distributions by only considering the contribution from the
nucleon pole. We can see that the angular distributions are symmetry
between the backward and forward angles. Comparing
Fig.~\ref{figangle1} with Fig.~\ref{figangle1n}, we see that, there
is a big difference between the full contribution and the only
nucleon contribution. Our model predictions may be tested by the
future experiments.

Note that the exchanged nuclear resonances in Fig.~\ref{cro} are far
off mass shell, and the form factors for exchanged nuclear
resonances here should be different with those that have been used
for the $e^-e^+\to \psi(3770) \to p \bar{p} \pi^0$ reaction. We know
that the form factors can be directly related to the hadron
structure. However, the question of hadron structure is still very
open, we have to adjust the form factor to fit the experimental
data, and the hadronic form factors are commonly used
phenomenologically~\cite{Feuster:1997pq,Haberzettl:1998eq,Yoshimoto:1999dr,Oh:2000zi}.
The effects of these form factors could substantially change the
predicted cross sections. Because of the lack of the available
experimental measurements, we can not determine the form factors
without ambiguities. In the present work, we take the same form
factors for both $\bar{p} p \to \psi(3770) \pi^0$ reaction and $e^+
e^- \to \psi(3770) \to \bar{p} p \pi^0$ reaction.

\section{The implication for $\psi(3686)\to p\bar{p}\pi^0$ and $p\bar p\to \psi(3686)\pi^0$}\label{sec4}

For the process $\psi(3686)\to p\bar{p}\pi^0$, we first determine
the coupling constant $g_{\psi(3686)NN}$, i.e., by using the
Lagrangian in Eq.~(\ref{vnn}), $g_{\psi(3686)NN}$ can be fitted
through the process $\psi(3686)\rightarrow p\bar{p}$. With the
experimental value \cite{Agashe:2014kda} $B(\psi(3686)\rightarrow
p\bar{p}$)$=2.8\times10^{-4}$, $g_{\psi(3686)NN}$ is determined to
be
\begin{eqnarray} \label{g3686nn}
g_{\psi(3686)NN} = 9.4 \times 10^{-4},
\end{eqnarray}
which is consistent with that given in Ref.~\cite{Barnes:2006ck}.

\begin{figure}[htpb]
\centering
\includegraphics[scale=0.3]{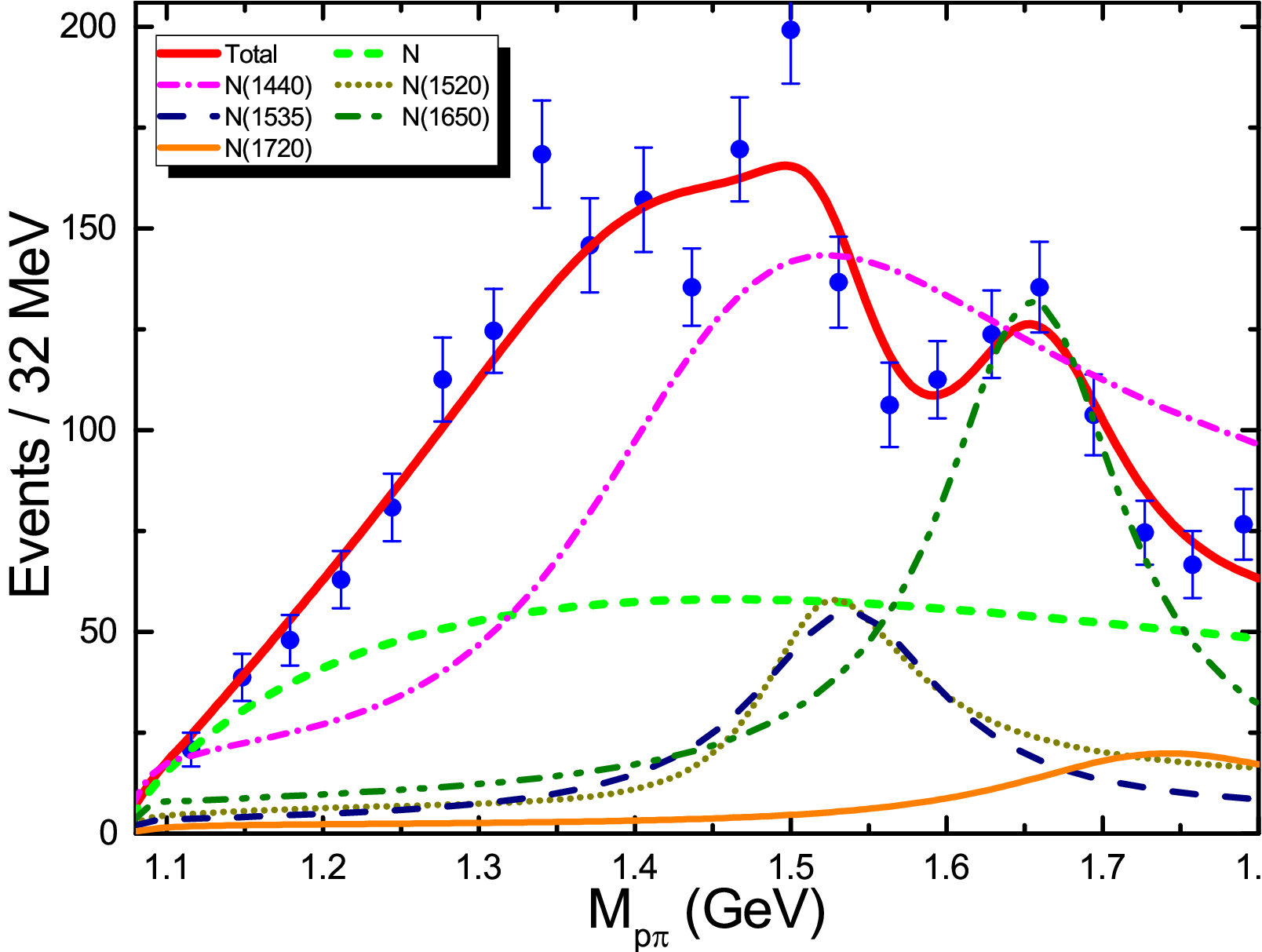}
\caption{(color online). The fitted $p \pi$ invariant mass spectrum
of the process of $\psi(3686) \to p\bar{p}\pi^0$. The dashed green curve
stands for the contribution of the nucleon pole, the solid red
line stands for the full contributions, and
other lines show the contributions from different $N^*$
resonances. The experiment data are
taken from Ref.~\cite{Ablikim:2012zk}.}\label{figdec}
\end{figure}

In Ref.~\cite{Ablikim:2012zk}, BESIII released the $p\pi$ invariant
mass spectrum of the process $\psi(3686) \to p\bar{p}\pi^0$ and
decay width $\Gamma(\psi(3686)\rightarrow
p\bar{p}\pi^0)=(1.65\pm0.03\pm0.15)\times10^{-5}$. Similar to the
case of $\psi(3770)$, we fit five coupling constants $g_{N^*}$, five
phase angles and a cut off parameter $\beta$ to the experimental
data. The fitted results are shown in Fig.~\ref{figdec}. Here, one
gets $\chi^2/d.o.f=2.90$ and $\beta=3.28 \pm 2.23$, while the fitted
coupling constants $g_{N^*}$ and phase angles are listed in
Table~\ref{ngphi2}.

\renewcommand{\arraystretch}{1.5}
\begin{table}[htbp]
    \caption{Fitted coupling constants $g_{N^*}$ and phase angles
        $\phi_{N^*}$ in the process $\psi(3686) \to p\bar{p}\pi^0$, where $g_{N^*} = g_{\psi(3686)NN^*}g_{\pi N N^*}$}.
    \label{ngphi2}
    \begin{center}
        \begin{tabular}{lcc} \toprule[0.6pt] \toprule[0.6pt]
            $N^*$     & $g_{N^*}\,(\times10^{-3})$     & $\phi_{N^*}\,(\textnormal{rad})$    \\ \midrule[0.2pt]
            $N(1440)$ & $5.10 \pm 0.86$  & $3.40 \pm 0.22$     \\ 
            $N(1520)$ & $2.27 \pm 0.39$  & $4.96 \pm 1.10$   \\ 
            $N(1535)$ & $0.51 \pm 0.36$ &  $0.75 \pm 0.64$  \\ 
            $N(1650)$ & $0.76 \pm 0.19$ &  $5.35 \pm 0.92$  \\ 
            $N(1720)$ & $0.98 \pm 0.42$ &  $1.77 \pm 0.99$  \\ \bottomrule[0.7pt]\bottomrule[0.7pt]
        \end{tabular}
    \end{center}
\end{table}

In Fig.~\ref{figdec}, the dashed curve stands for the contribution
of the nucleon pole, the solid line stands for the full
contributions, and
other lines show the contributions from different $N^*$
resonances. We see that we can describe the experimental data
fairly well. Furthermore, we find that the peak between 1.6 GeV and
1.7 GeV mainly comes from the contribution of $N(1650)$.

There also exist quiet obvious interfering effects between different
$N^*$ contributions in Fig.~\ref{figdec}. Close to $M_{p \pi} = 1.6$
GeV, comparing the $N(1440)$ contribution to the total contribution,
one can see the $N(1440)$ contribution is "digged out" a valley by
other $N^*$ contributions. In the region of $M_{p \pi} > 1.7$ GeV,
the total contribution is smaller than the $N(1440)$ contribution,
i.e., the total contribution is suppressed by interfering terms. So,
from Fig.~\ref{figdec2} and Fig.~\ref{figdec}, one can see how
important the interference effect is. We will not be able to get a
good fit without interfering terms and arbitrary phase angles.

Additionally, we also calculated the branch fractions of
$\psi(3686)\rightarrow (N^*\bar{p}+c.c.)\rightarrow p\bar{p}\pi^0$
from individual intermediate $N^*$ (or $p$) state, with the fitted
coupling constants listed in the Table \ref{ngphi2}. Our results are
shown in Table~\ref{bf}. The errors of our theoretical results are
obtained from the errors of those fitted coupling constants of
$g_{N^*}$. We also notice that in Ref.~\cite{Ablikim:2012zk} BESIII
also extracted the corresponding branching fractions without
considering the interference of different intermediate $N^*$ (or
$p$) states, which is different from the treatment in the present
work. Thus, in Table \ref{bf} we further compare our result with the
experimental results~\cite{Ablikim:2012zk}, we see that our results
are in agreement within errors with that given in
Ref.~\cite{Ablikim:2012zk}.

\setlength{\tabcolsep}{12pt}
\renewcommand{\arraystretch}{1.5}
\begin{table}[htbp]
    \caption{The calculated branching fractions $\psi(3686)
        \rightarrow p\bar{p}\pi^0$ if considering individual intermediate $N^*$
        (or $N$) contribution, and the comparison with the experiment values
        of Ref.~\cite{Ablikim:2012zk}. Here, all values are in the unit of
        $10^{-5}$.}
    \label{bf}
    \begin{center}
        \begin{tabular}{lcc} \toprule[0.7pt]\toprule[0.7pt]
            & Our results & The results in Ref. \cite{Ablikim:2012zk} \\ \midrule[0.2pt]
            $N$       & $7.5$ & $6.42^{+0.20+1.78}_{-0.20-1.28}$ \\ 
            $N(1440)$ &$14 \pm 4.5$ & $3.58^{+0.25+1.59}_{-0.25-0.84}$ \\
            $N(1520)$ &$2.8 \pm 0.8$  & $0.64^{+0.05+0.22}_{-0.05-0.17}$ \\
            $N(1535)$ &$2.1 \pm 3.0$  & $2.47^{+0.28+0.99}_{-0.28-0.97}$ \\
            $N(1650)$ &$4.9 \pm 2.5$  & $3.76^{+0.28+1.37}_{-0.28-1.66}$ \\
            $N(1720)$ &$1.3 \pm 1.0$  & $1.79^{+0.10+0.24}_{-0.10-0.71}$ \\\bottomrule[0.7pt]\bottomrule[0.7pt]
        \end{tabular}
    \end{center}
\end{table}

\begin{figure}[htpb]
    \centering
    \includegraphics[scale=0.27]{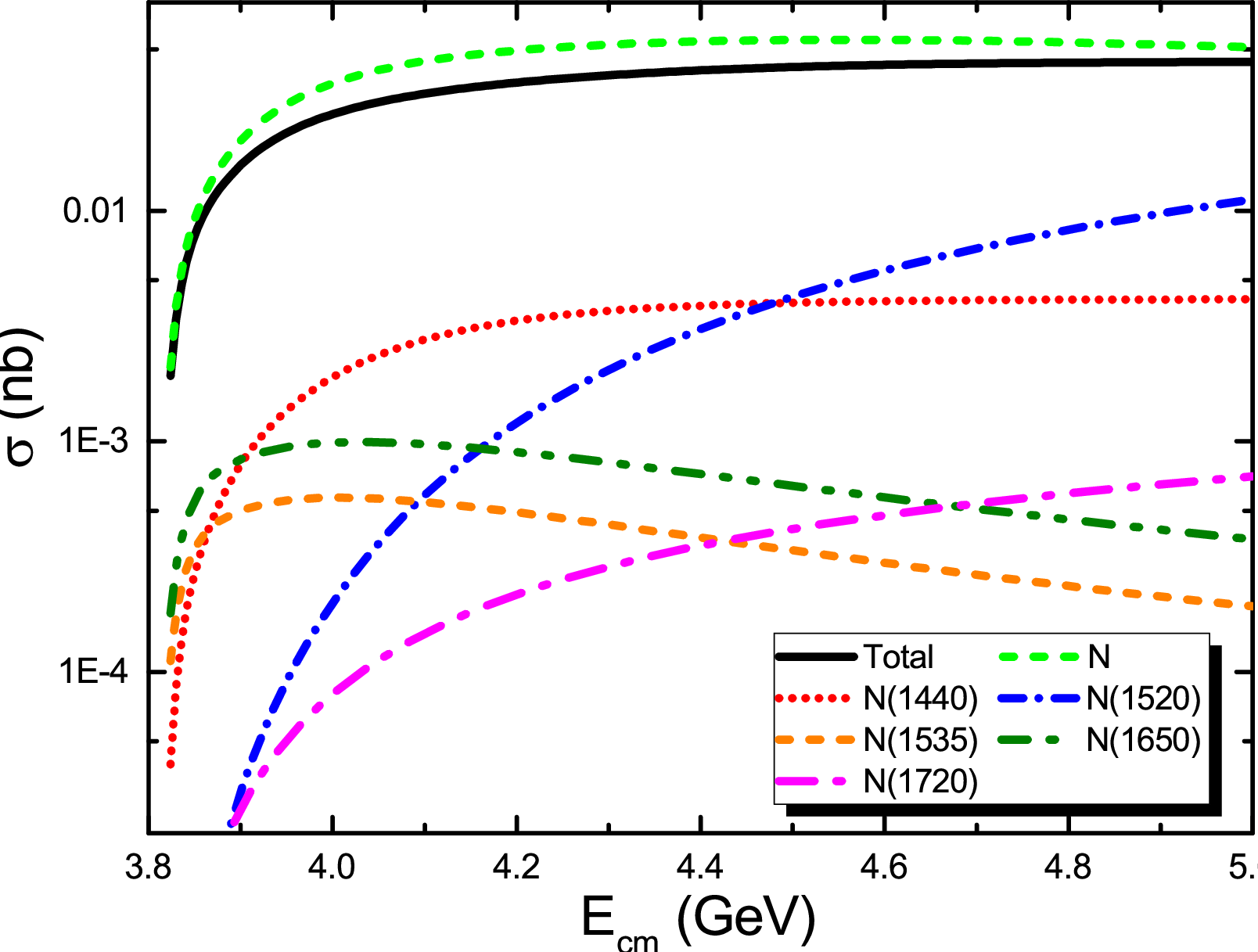}
    \caption{(color online). The cross section of the
        process $p \bar p \rightarrow \pi^0  \psi(3686)$. The black line is
        total result, and other lines shows
        the $N^*$ contribution.}\label{figcros11}
\end{figure}

With these fitted parameters, we calculate the cross section of the
process $p \bar p \rightarrow \pi^0 \psi(3686)$ with cross symmetry.
The results are shown in Fig.~\ref{figcros11}. One can see that the
nucleon pole contribution is predominant in the whole energy region,
while the contributions from other $N^*$ states are small. In the
higher energy region, the nucleon pole contribution is starting to
decrease, while the full contribution increases slowly, this
behavior resembles the process $p \bar p \rightarrow \pi^0
\psi(3770)$. Furthermore, it is noticed that, the discrepancy
between the total result and the nucleon contribution is smaller
than the case of $p \bar p \to \pi^0 \psi(3770)$.

\begin{figure}[htpb]
    \centering
    \includegraphics[scale=0.27]{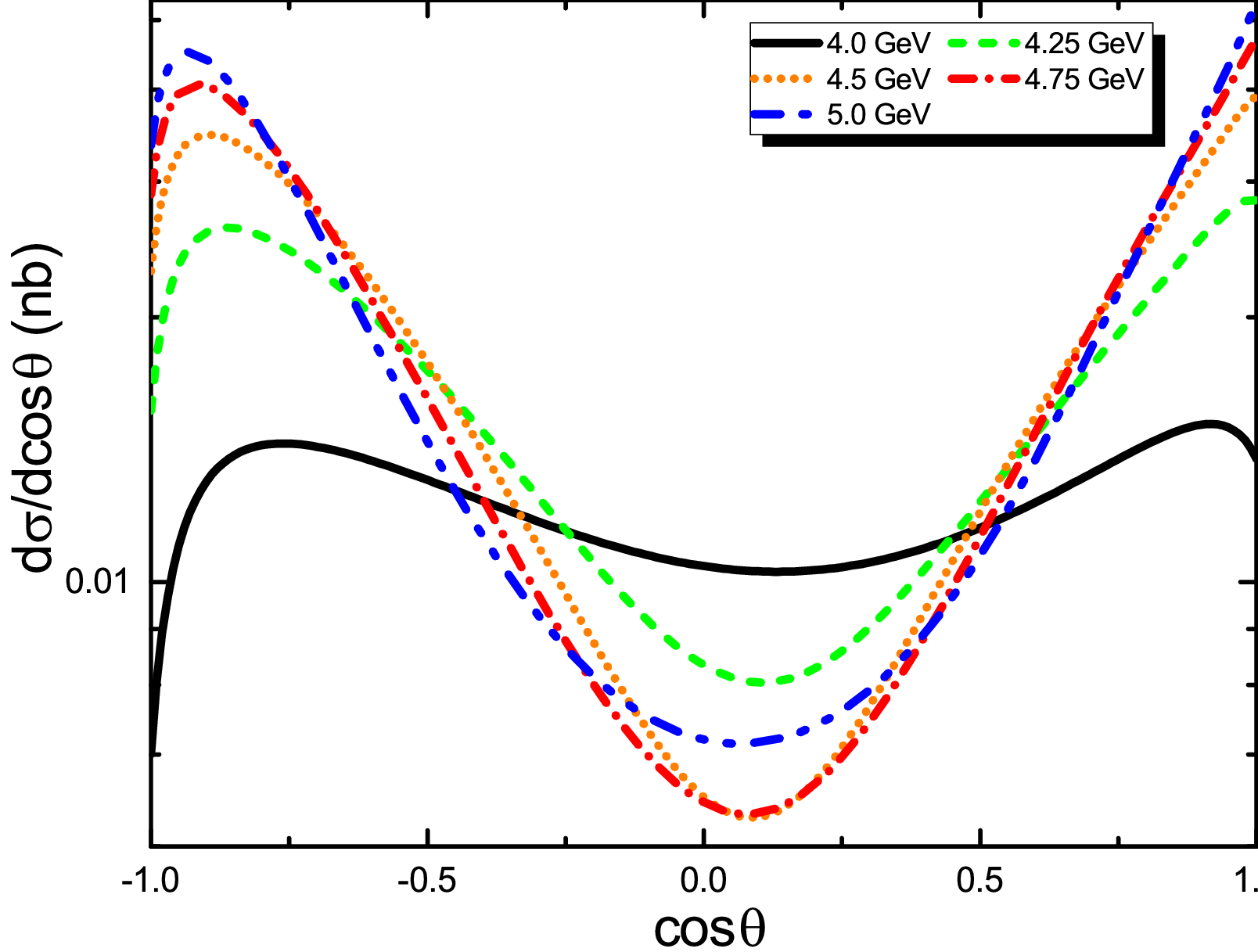}
    \caption{(color online). the angular distribution of the process $p
        \bar p \rightarrow \pi^0  \psi(3686)$. Each line shows the different
        c.m. energy.}\label{figangle2}
\end{figure}

\begin{figure}[htpb]
    \centering
    \includegraphics[scale=0.27]{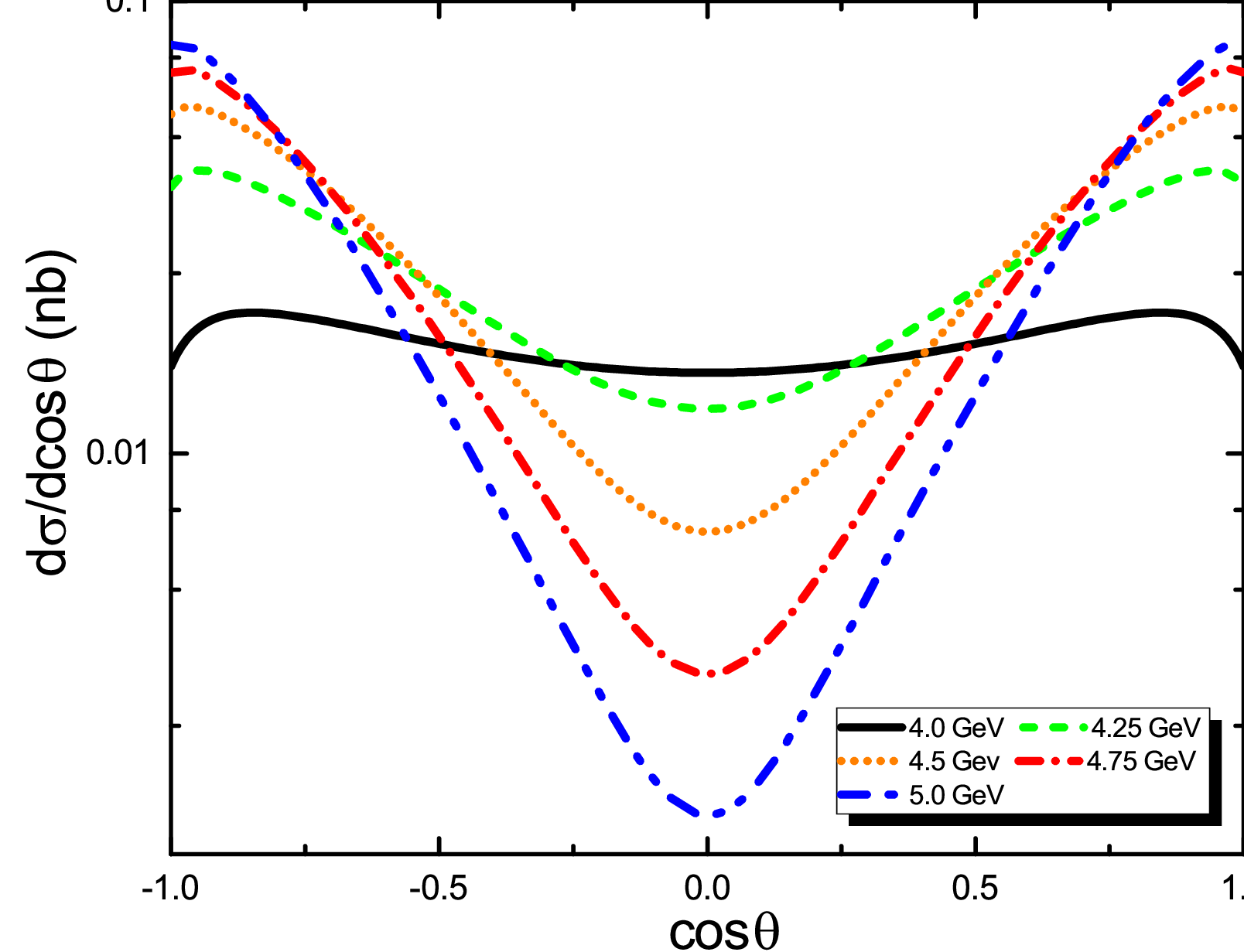}
    \caption{(color online). the angular distribution of the process $p
        \bar p \rightarrow \pi^0  \psi(3686)$ only considering the nucleon
        contribution. Each line shows the different c.m.
        energy.}\label{figangle2n}
\end{figure}

Finally, we show the angular distributions of the process $p \bar p
\rightarrow \pi^0  \psi(3686)$ in Figs.~\ref{figangle2} and
\ref{figangle2n}. Similar to Fig. \ref{figangle1}, there is a peak
in backward angle and a valley close to $\cos\theta=0$. Comparing to
the angular distribution with the nucleon contribution in Fig.
\ref{figangle2n}, there exits obvious difference, since the nucleon
contribution only is symmetry while total contribution is asymmetry.

\section{discussion and conclusion}\label{sec5}

We have studied the $e^+e^-  \to p \bar{p} \pi^0$ at $3.773$ GeV c.m. energy and $p
\bar p \to \pi^0  \psi(3770)$ reaction within an effective
lagrangian approach. The $e^+e^-   \to p \bar{p} \pi^0$ process is a
good platform to study excited $N^*$ nucleon resonances. We consider
contributions from nucleon pole and five well established $N^*$
states. First, we perform a $\chi^2$-fit to the experimental data on
the mass distribution of the $e^+e^-  \to p \bar{p} \pi^0$ ,
from where we obtain the couplings of $\psi(3770)$ to these $N^*$
states. It is shown that we can describe the experimental data quite
well. In particular, the two bumps around $1.5$ and $1.7$ GeV can be
well reproduced. We also find that the contribution of the nucleon
pole is small comparing to the background contribution, and there exists large cancellation in some $M_{\bar{p}\pi}$ regions.

Second, based on our results of the $e^+e^- \to p \bar{p} \pi^0$, we study the $p \bar p \to \pi^0 \psi(3770)$ reaction with
cross symmetry. We evaluate the total and differential cross
sections of the $p \bar p \to \pi^0 \psi(3770)$ reaction. The nucleon pole gives largest contribution to the $p
\bar p \to \pi^0 \psi(3770)$ reaction within a wide range. However,
the interference terms between nucleon pole and the other nucleon
resonance affects significantly and could change the angle
distributions clearly. Our studies provide valuable information to
future experimental exploration the $\psi(3770) \pi^0$ production
through the $p \bar p$ interaction.

Additionally, we also study the $\psi(3686)$ production through the
process $p \bar p \to \pi^0 \psi(3686)$. Similarly to the case of
$e^+e^- \to \psi(3770) \to p \bar{p} \pi^0$, we study firstly the
decay process of $\psi(3686) \to p \bar{p} \pi^0$ to extract the
parameters we needed. Then we study the $p \bar p \to \pi^0
\psi(3686)$ reaction. We find that the contribution from the nucleon
pole is dominant, while the angular distributions show a quite
discrepancy induced by the $N^*$ states.

We hope and expect that future experiments at
$\overline{\mbox{P}}$ANDA will provide a test to our model and give
more constraints on our theoretical study.

\section*{Acknowledgments}

This project is supported by the National Natural Science Foundation
of China under Grants No. 11222547, No. 11175073, and No. 11475227,
the Ministry of Education of China (SRFDP under Grant No.
2012021111000), and the Fok Ying Tung Education Foundation (Grant
No. 131006). This work is also supported by the Open Project Program
of State Key Laboratory of Theoretical Physics, Institute of
Theoretical Physics, Chinese Academy of Sciences, China
(No.Y5KF151CJ1).

\section*{Appendix: Scattering amplitudes of the subprocess $\psi(3770) \to  p\bar{p}\pi^0$ and the process $p \bar p \rightarrow \pi^0  \psi(3770)$.}

The tree level diagrams of the subprocess $\psi(3770) \to p \bar{p}\pi^0$ is
depicted in part of Fig.~\ref{dec} (a). According to the feynmann diagrams shown
in Fig.~\ref{dec}, the scattering amplitudes $\mathcal{M}_{J^P}$
with a exchanged $N^*(J^P)$ (including $N$) are given by:
\begin{widetext}
\begin{eqnarray}
\mathcal{M}_{\frac{1}{2}^+} &=& \frac{g_{\pi NP_{11}}}{2m_N}
g_{VNP_{11}} \epsilon^\mu(p_1) \overline{u}(p_2) \left[\gamma_\mu
\frac{-\slashed{p_t}+m_{N^*}}{t-m_{N^*}^2} \gamma_5 (i\slashed{p_4})
F(t) + \gamma_5 (i\slashed{p_4}) \frac{\slashed{p_u} +
m_{N^*}}{u-m_{N^*}^2} \gamma_\mu F(u) \right] v(p_3),  \label{amp0}
\\
\mathcal{M}_{\frac{1}{2}^-} &=& g_{\pi NS_{11}} g_{VNS_{11}}
\epsilon^\mu(p_1) \overline{u}(p_2) \left[\gamma_5 \gamma_\mu
\frac{-\slashed{p_t}+m_{N^*}}{t-m_{N^*}^2} F(t) +
\frac{\slashed{p_u} +m_{N^*}}{u-m_{N^*}^2} \gamma_5 \gamma_\mu F(u)
\right] v(p_3), \\
\mathcal{M}_{\frac{3}{2}^+} &=& \frac{g_{\pi NP_{13}}}{m_N} i
g_{VNP_{13}} \epsilon^\mu(p_1) \overline{u}(p_2) \left[\gamma_5
\frac{-\slashed{p_t}+m_{N^*}}{t-m_{N^*}^2} G_{\mu\nu}(-p_t) (i p^\nu_4)
F(t) + (i p^\nu_4) \frac{\slashed{p_u} +m_{N^*}}{u-m_{N^*}^2}
G_{\nu\mu}(p_u) \gamma_5 F(u) \right] v(p_3), \\
\mathcal{M}_{\frac{3}{2}^-}&=& \frac{g_{\pi ND_{13}}}{m_N^2}
g_{VND_{13}} \epsilon^\mu(p_1) \overline{u}(p_2)
\left[\frac{-\slashed{p_t}+m_{N^*}}{t-m_{N^*}^2} G_{\mu\nu}(-p_t)
\gamma_5 (i \slashed{p_4}) (i p^\nu_4) F(t) + \gamma_5 (i \slashed{p_4}) (i p^\nu_4)
\frac{\slashed{p_u} +m_{N^*}}{u-m_{N^*}^2} G_{\nu\mu}(p_u) F(u)
\right] v(p_3)\label{amp1}
\end{eqnarray}
with $t = p^2_t = (p_1-p_2)^2$ and $u = p^2_u = (p_1 - p_3)^2$, and
$G_{\mu\nu}$ is
\begin{eqnarray}
G_{\mu\nu}(p) = (-g_{\mu\nu}+\frac{1}{3} \gamma_\mu \gamma_\nu +
\frac{1}{3m_{N^*}}(\gamma_\mu p_\nu- \gamma_\nu p_\mu) + \frac{2}{3}
\frac{p_\mu p_\nu}{m_{N^*}^2}).
\end{eqnarray}

For $\mathcal{M}_{N^*(N)}^\alpha$ in Eq.~(\ref{mpppi}), just drop
the polarization vector $\epsilon^\mu(p_1)$.

For $p \bar p \to \pi^0  \psi(3770)$ reaction, the scattering
amplitudes can be easily obtained just applying the substitution to
Eqs.~(\ref{amp0}-\ref{amp1}):
\begin{eqnarray}
p_1\rightarrow-p_3, \quad p_2\rightarrow-p_2, \quad p_3\rightarrow-p_1, \quad p_t\rightarrow -p_u.
\end{eqnarray}

The amplitude of the process $\psi(3686) \to  p\bar{p}\pi^0$ and $p \bar p \rightarrow \pi^0  \psi(3686)$ is exactly the same.
\end{widetext}


\begin{thebibliography}{99}

\bibitem{Lutz:2009ff}
M.~F.~M.~Lutz {\it et al.}  [PANDA Collaboration],
arXiv:0903.3905 [hep-ex].

\bibitem{Gaillard:1982zm}
  M.~K.~Gaillard, L.~Maiani and R.~Petronzio,
  Phys.\ Lett.\ B {\bf 110}, 489 (1982).

\bibitem{Lundborg:2005am}
  A.~Lundborg, T.~Barnes and U.~Wiedner,
  Phys.\ Rev.\ D {\bf 73}, 096003 (2006).

\bibitem{Barnes:2006ck}
  T.~Barnes and X.~Li,
  Phys.\ Rev.\ D {\bf 75}, 054018 (2007).

\bibitem{Barnes:2007ub}
  T.~Barnes, X.~Li and W.~Roberts,
  Phys.\ Rev.\ D {\bf 77}, 056001 (2008).

\bibitem{Barnes:2010yb}
T.~Barnes, X.~Li and W.~Roberts,
Phys.\ Rev.\ D {\bf 81}, 034025 (2010). 

\bibitem{Lin:2012ru}
  Q.~Y.~Lin, H.~S.~Xu and X.~Liu,
  Phys.\ Rev.\ D {\bf 86}, 034007 (2012).

\bibitem{Pire:2013jva}
  B.~Pire, K.~Semenov-Tian-Shansky and L.~Szymanowski,
  Phys.\ Lett.\ B {\bf 724}, 99 (2013).

\bibitem{Wiele:2013vla}
  J.~Van de Wiele and S.~Ong,
  Eur.\ Phys.\ J.\ C {\bf 73}, 2640 (2013).

\bibitem{Ablikim:2014kxa}
M.~Ablikim {\it et al.}  [BESIII Collaboration],
Phys.\ Rev.\ D {\bf 90}, 032007 (2014). 

\bibitem{Klempt:2009pi}
  E.~Klempt and J.~M.~Richard,
  Rev.\ Mod.\ Phys.\  {\bf 82}, 1095 (2010).

\bibitem{Ablikim:2004ug}
  M.~Ablikim {\it et al.}  [BES Collaboration],
  Phys.\ Rev.\ Lett.\  {\bf 97}, 062001 (2006).


\bibitem{Barnes:2011pm}
T.~Barnes,
Int.\ J.\ Mod.\ Phys.\ Conf.\ Ser.\  {\bf 02}, 193 (2011).

\bibitem{Ablikim:2012zk}
M.~Ablikim {\it et al.}  [BESIII Collaboration],
Phys.\ Rev.\ Lett.\  {\bf 110}, 022001 (2013).


\bibitem{Agashe:2014kda}
K.~A.~Olive {\it et al.}  [Particle Data Group Collaboration],
Chin.\ Phys.\ C {\bf 38}, 090001 (2014).

\bibitem{Lin:2013mka}
Q.~Y.~Lin, X.~Liu and H.~S.~Xu,
Phys.\ Rev.\ D {\bf 88}, 114009 (2013).

\bibitem{Tsushima:1996xc}
K.~Tsushima, A.~Sibirtsev and A.~W.~Thomas,
Phys.\ Lett.\ B {\bf 390}, 29 (1997). 

\bibitem{Tsushima:1998jz}
K.~Tsushima, A.~Sibirtsev, A.~W.~Thomas and G.~Q.~Li,
Phys.\ Rev.\ C {\bf 59} (1999) 369 [Phys.\ Rev.\ C {\bf 61} (2000)
029903]. 

\bibitem{Zou:2002yy}
B.~S.~Zou and F.~Hussain,
Phys.\ Rev.\ C {\bf 67}, 015204 (2003). 

\bibitem{Ouyang:2009kv}
Z.~Ouyang, J.~J.~Xie, B.~S.~Zou and H.~S.~Xu,
Int.\ J.\ Mod.\ Phys.\ E {\bf 18}, 281 (2009). 

\bibitem{Wu:2009md}
J.~J.~Wu, Z.~Ouyang and B.~S.~Zou,
Phys.\ Rev.\ C {\bf 80}, 045211 (2009). 

\bibitem{Cao:2010km}
X.~Cao, B.~S.~Zou and H.~S.~Xu,
Phys.\ Rev.\ C {\bf 81}, 065201 (2010). 

\bibitem{Cao:2010ji}
X.~Cao, B.~S.~Zou and H.~S.~Xu,
Nucl.\ Phys.\ A {\bf 861} (2011) 23. 

\bibitem{Huang:2005js}
S.~Z.~Huang, P.~F.~Zhang, T.~N.~Ruan, Y.~C.~Zhu and Z.~P.~Zheng,
Eur.\ Phys.\ J.\ C {\bf 42}, 375 (2005).


\bibitem{Feuster:1997pq}
T.~Feuster and U.~Mosel,
Phys.\ Rev.\ C {\bf 58}, 457 (1998). 

\bibitem{Haberzettl:1998eq}
H.~Haberzettl, C.~Bennhold, T.~Mart and T.~Feuster,
Phys.\ Rev.\ C {\bf 58} (1998) 40. 

\bibitem{Yoshimoto:1999dr}
T.~Yoshimoto, T.~Sato, M.~Arima and T.~S.~H.~Lee,
Phys.\ Rev.\ C {\bf 61}, 065203 (2000). 

\bibitem{Oh:2000zi}
Y.~s.~Oh, A.~I.~Titov and T.~S.~H.~Lee,
Phys.\ Rev.\ C {\bf 63}, 025201 (2001). 

\bibitem{Chen:2010nv}
D.~Y.~Chen, J.~He and X.~Liu,
Phys.\ Rev.\ D {\bf 83}, 054021 (2011)
[arXiv:1012.5362 [hep-ph]].

\bibitem{Xie:2015zga}
J.~J.~Xie, Y.~B.~Dong and X.~Cao,
Phys.\ Rev.\ D {\bf 92}, 034029 (2015).


\end{thebibliography}
\end{document}